\documentstyle[12pt]{article}
\title{Renormalization of vacuum energy in linearized quantum gravity}
\author{Hrvoje Nikoli\'c \\
Theoretical Physics Division, Rudjer Bo\v{s}kovi\'{c} Institute, \\
P.O.B. 180, HR-10002 Zagreb, Croatia \\
{\normalsize hrvoje@thphys.irb.hr} \\
\makebox[1in]{} \\
}
\date{\today}
\begin{document}
\maketitle
\begin{abstract}
In linearized quantum gravity, a shift of the average energy-momentum 
can be compensated by a shift of the average gravitational field.
This allows a renormalization scheme that naturally removes the
contribution of quantum vacuum fluctuations to the cosmological constant, 
solving the old cosmological-constant problem for weak gravitational fields.
\end{abstract}
\vspace*{0.5cm}
\vspace*{0.9cm}

As is well known, the average energy of a quantum
field diverges in any quantum state, including the vacuum. 
In non-gravitational physics the zero-energy point can be chosen
arbitrarily, so energy can be renormalized by a simple
shift of energy chosen such that the energy of the vacuum
is equal to zero. However, in gravitational physics
the energy is the source of the gravitational field, so 
the zero-energy point is not arbitrary. Consequently, 
the quantum vacuum energy contributes to the cosmological
constant, by a contribution 120 orders of magnitude larger 
than the measured one. This problem is known as the 
cosmological-constant problem \cite{weinberg,nobb,sahni,carroll,padm}. 
In the old
formulation of the problem \cite{weinberg,nobb} one would like
to find a theoretical mechanism that
makes this vacuum contribution to the cosmological constant
vanishing. In the new, more ambitious, formulation of the
problem \cite{sahni,carroll,padm} one would like to explain
why the sum of all possible contributions to the cosmological
constant, including that of the vacuum energy, is of the same
order of magnitude as the matter density of the universe.

The cosmological-constant problem usually appears at the level
of semi-classical gravity, in which matter is quantized but gravity is not.
It is very likely that the solution of the problem requires a fully
quantized gravity. A renormalization of the energy-momentum could be
compensated by a renormalization of the gravitational field,
which could make the subtraction of the vacuum energy-momentum
consistent.
Unfortunately, a fully quantized gravity
is too difficult to deal with, so one is forced to use some
approximations. In this letter we deal with linearized quantum gravity
and show how in this context the old cosmological-constant problem
can be solved in a very simple way. It also represents a very generic solution
of the problem, in the sense that it does not depend on technical details 
specifying how exactly gravity is quantized.

In an appropriate system of units, the classical Einstein equation can be written
as $G_{\mu\nu}=T_{\mu\nu}$. The left-hand side $G_{\mu\nu}$ is the Einstein tensor, which can be expressed as a nonlinear functional of the metric $g_{\mu\nu}$. The right-hand side $T_{\mu\nu}$ is the matter
energy-momentum tensor, which can be expressed as a nonlinear functional of the metric $g_{\mu\nu}$ and the matter fields generically denoted by $\phi$. 
To simplify the notation,
in the rest of the discussion we suppress the spacetime indices $\mu,\nu$.
This allows us to write the Einstein equation in a generic form as
\begin{equation}\label{in1}
G[g]=T[g,\phi] .
\end{equation}
The metric can be written as 
\begin{equation}\label{in2}
g=\eta+h ,
\end{equation}
where $\eta$ is the flat Minkowski metric and $h$ is the dynamical gravitational
field. Here $\eta$ is treated as a fixed metric, so (\ref{in1}) can be
written as
\begin{equation}\label{in3}
G[h]=T[h,\phi] ,
\end{equation} 
where now $G$ and $T$ are some new functionals
highly nonlinear in $h$ (see, e.g., \cite{feyn}). 

In quantum gravity, the fields $h$ and $\phi$ are promoted to the 
operators $\hat{h}$ and $\hat{\phi}$. Thus, with an appropriate 
operator ordering, the quantum fields in the Heisenberg picture satisfy the
operator Einstein equation
\begin{equation}\label{in4}
G[\hat{h}]=T[\hat{h},\hat{\phi}] .
\end{equation}
Hence, for any {\em physical} quantum state $|\psi\rangle$ we have
\begin{equation}\label{in5}
\langle\psi| G[\hat{h}] |\psi\rangle = 
\langle\psi| T[\hat{h},\hat{\phi}] |\psi\rangle .
\end{equation}
In particular, we choose one special fiducial state $|\Omega\rangle$ 
(to be specified below), so we have
\begin{equation}\label{in6}
\langle\Omega| G[\hat{h}] |\Omega\rangle = 
\langle\Omega| T[\hat{h},\hat{\phi}] |\Omega\rangle .
\end{equation}
With a generic ordering of operators,
the right-hand sides of (\ref{in5}) and (\ref{in6}) are
divergent, and hence so are the left-hand sides.
Still, we can subtract these two equations, leading to
\begin{equation}\label{in7}
\langle\psi| G[\hat{h}] |\psi\rangle
- \langle\Omega| G[\hat{h}] |\Omega\rangle = T^{(\psi)}, 
\end{equation} 
where 
\begin{equation}\label{in8}
T^{(\psi)} \equiv \langle\psi| T[\hat{h},\hat{\phi}] |\psi\rangle 
- \langle\Omega| T[\hat{h},\hat{\phi}] |\Omega\rangle .
\end{equation} 
We see that (\ref{in8}) looks just as the average energy-momentum tensor
in the state $|\psi\rangle$, renormalized by subtraction of the 
average energy-momentum tensor in the fiducial state $|\Omega\rangle$.
Thus, if $|\Omega\rangle$ was the matter vacuum, the right-hand side 
of (\ref{in7}) would look just like the right-hand side of the 
semiclassical Einstein-equation with a shifted energy-momentum,
such that the vacuum energy-momentum vanishes. If the left-hand side
of (\ref{in7}) could be interpreted as the left-hand side
of the semiclassical Einstein equation, this would represent a 
solution of the old cosmological-constant problem, because  
the vacuum energy-momentum would be removed by a shift of the zero-energy point,
without ignoring the fact that the vacuum energy-momentum is also 
a source for the gravitational field.
Unfortunately, the problem is that
the left-hand side of the semiclassical Einstein equation
should have a form $G[h]$, implying that the left-hand side of
(\ref{in7}) {\em cannot} be interpreted as the left-hand side
of the semiclassical Einstein equation. 
Thus, it is not yet clear how the simple manipulations above can help 
in solving the cosmological-constant problem.

The situation significantly improves in linearized gravity.
The functional $G[h]$ can be expanded in powers of $h$ as
\begin{equation}\label{in9}
G[h] \simeq Lh + {\cal O}(h^2) ,
\end{equation}
where $L$ is a linear operator (essentially a 
second-order derivative operator not depending on $h$)
acting on $h(x)$. Linearized gravity is an approximation in which
${\cal O}(h^2)$ in (\ref{in9}) is neglected. It is a good 
approximation for weak gravitational fields. In the lowest order
in $h$ we can also write $T[h,\phi]\simeq T[\phi]$, i.e., calculate
the energy-momentum tensor with the flat Minkowski background metric $\eta$.
Therefore, in the lowest order in $\hat{h}$, (\ref{in7}) can be
written as
\begin{equation}\label{in10}
Lh^{(\psi)} = T^{(\psi)}, 
\end{equation} 
where
\begin{equation}\label{in8.1}
T^{(\psi)} = \langle\psi| T[\hat{\phi}] |\psi\rangle 
- \langle\Omega| T[\hat{\phi}] |\Omega\rangle ,
\end{equation} 
\begin{equation}\label{in11}
h^{(\psi)} \equiv \langle\psi| \hat{h} |\psi\rangle 
- \langle\Omega| \hat{h} |\Omega\rangle .
\end{equation}
We see that (\ref{in10}) looks exactly as a linearized 
semiclassical Einstein-equation, in which the ``classical" 
gravitational field
$h^{(\psi)}$ is actually the average gravitational field
renormalized by subtraction of the 
average gravitational field in the fiducial state $|\Omega\rangle$.

So far, we have not fixed the fiducial state $|\Omega\rangle$.
We fix it by requiring that the vacuum energy-momentum
$T^{(0)}$ has the cosmological-constant form. In the lowest
order in $h$, this means
\begin{equation}\label{in12}
T^{(0)}=\Lambda \eta ,
\end{equation}
where $\Lambda$ is an {\em unspecified} scalar. Eq.~(\ref{in8.1})
then implies
\begin{equation}\label{in13}
\langle 0_{\phi} | T[\hat{\phi}] | 0_{\phi} \rangle 
- \langle\Omega| T[\hat{\phi}] |\Omega\rangle = \Lambda \eta ,
\end{equation}
where $| 0_{\phi} \rangle$ is the matter vacuum. But (\ref{in13}) is 
only possible if $|\Omega\rangle$ is also proportional to 
a Lorentz invariant matter state, that is, to the matter vacuum.
Hence, assuming that the matter vacuum $|0_{\phi}\rangle$ is unique,
$|\Omega\rangle$ must be of the form
\begin{equation}\label{in14}
|\Omega\rangle = |\Omega_h\rangle \otimes | 0_{\phi} \rangle ,
\end{equation}
where $|\Omega_h\rangle$ is some gravitational state. 
This implies that $T^{(0)}=0$, 
i.e., that the contribution of the quantum fluctuations
to the cosmological constant in (\ref{in12}) is
\begin{equation}\label{0}
\Lambda=0 .
\end{equation}
We also note that (\ref{in8.1}) is finite and that,
due to (\ref{in14}), it corresponds to the usual normal ordering 
of matter field operators. This is so owing to the fact that the background
metric $\eta$ is chosen to be the flat Minkowski one.
(In $\eta$ was chosen to be a curved background, then (\ref{in8.1}) would
not necessarily be finite \cite{bd}.) 
Thus, {\em the quantum fluctuations of the vacuum do not contribute
to the cosmological constant} in linearized quantum gravity. 

Further, our renormalization scheme removes even a 
contribution from a bare cosmological constant. To see this, 
assume that the classical
action contains a cosmological term 
proportional to $\int d^4x \sqrt{|g|} \Lambda_B$, 
where $\Lambda_B$ is a constant representing the bare cosmological constant.
The right-hand side of (\ref{in1}) attains an additional term
$\Lambda_B g$. In the lowest order in $h$ this becomes
$\Lambda_B \eta$, which in quantum theory does {\em not} become
an operator. Consequently, as 
$\langle\psi|\psi\rangle = \langle\Omega|\Omega\rangle =1$,
the right-hand sides of (\ref{in5})
and (\ref{in6}) attain the same additional term $\Lambda_B \eta$, 
which cancel each other after the subtraction in (\ref{in7}).

Our results show that a natural value of the cosmological constant
in linearized quantum gravity is zero. This
solves the old cosmological-constant problem in linearized
quantum gravity. But what about the new cosmological-constant problem?
Although our approach does not solve it,  
our result (\ref{0}) is not in contradiction
with the fact that the measured cosmological constant is not really zero.
Namely, although $T^{(0)}=0$, it does not imply that
$T^{(\psi)}$ cannot have a contribution of the form 
$\Lambda^{(\psi)}(x)\eta$, where $\Lambda^{(\psi)}(x)$ is some 
dynamical contribution to dark energy, depending on the state 
$|\psi\rangle$. Indeed, such a contribution, if exists, must depend
on the matter content of the universe described by the
state $|\psi\rangle$, which could explain why the measured
energy density of dark energy
is of the same order of magnitude as the measured matter density.
Thus, the observed cosmological constant could be explained dynamically
by a quantum version of a quintessence model.

Another mechanism that may induce an effective cosmological
constant is by allowing the existence of more than one vacuum state,
having different energies. The vacuum in the first term in 
(\ref{in13}) may be $|0_{\phi,2}\rangle$, while the vacuum in
(\ref{in14}) may be $|0_{\phi,1}\rangle$. Then the difference
on the right-hand side of (\ref{in13}) may be finite.
Such a mechanism could be relevant for the early cosmological
inflation.

A further contribution to an effective cosmological constant
could arise from nonlinear corrections to our
linearized quantum gravity. At this level such an idea may seem 
rather speculative, but results from loop quantum gravity
already suggest that the early inflation may be a consequence of strong
quantum gravitational fields interacting with generic matter \cite{boj}. 

To conclude, in linearized quantum gravity
a shift of the average energy-momentum, Eq.~(\ref{in8.1}),  
can be compensated by a shift of the average gravitational field, Eq.~(\ref{in11}).
This leads to a simple resolution of the old cosmological-constant problem
for weak gravitational fields. It also gives some hints on
a possible resolution of the new cosmological-constant problem as well.

\section*{Acknowledgements}

This work was supported by the Ministry of Science of the
Republic of Croatia under Contract No.~098-0982930-2864.

\end{document}